\documentclass[10pt,a4paper,aps,preprint,superscriptaddress,nofootinbib]{revtex4-1}

\pdfoutput=1

\usepackage{graphicx,color,verbatim,epsfig,subfigure}
\usepackage{slashed}
\usepackage{latexsym,amsfonts,amssymb,amsmath,makeidx,mathrsfs,multirow,bm,float}
\usepackage[makeroom]{cancel}
\usepackage[utf8]{inputenc}

\usepackage[colorlinks=true, pdfstartview=FitV, linkcolor=blue, citecolor=blue, urlcolor=blue]{hyperref}
\allowdisplaybreaks[4]

\def\beqn{\begin{eqnarray}}
\def\eeqn{\end{eqnarray}}
\def\beqs{\begin{subequations}}
\def\eeqs{\end{subequations}}
\def\beq{\begin{equation}}
\def\eeq{\end{equation}}
\def\ba{\begin{array}}
\def\ea{\end{array}}

\def\hf{\frac{1}{2}}
\def\[{\left[}
\def\]{\right]}
\def\({\left(}
\def\){\right)}

\def\TeV{\rm TeV}
\def\GeV{\rm GeV}

\def\fb{\rm fb}

\def\mL{\mathcal{L}}

\def\mO{\mathcal{O}}

\begin{document}

\title{The collider tests of a leptophilic scalar for the anomalous magnetic moments}

\author{Ning Chen}
\email{chenning$\_$symmetry@nankai.edu.cn}
\affiliation{
School of Physics, Nankai University, Tianjin 300071, China
}
\author{Bin Wang}
\email{wb@mail.nankai.edu.cn}
\affiliation{
School of Physics, Nankai University, Tianjin 300071, China
}
\author{Chang-Yuan Yao}
\email{yaocy@nankai.edu.cn}
\affiliation{
School of Physics, Nankai University, Tianjin 300071, China
}


\begin{abstract}
We study the anomalous muon and electron magnetic moments by introducing a scalar with CP-violating Yukawa couplings to the lepton sector. By fitting these two magnetic moments with the recent experimental measurements, we find that such a leptophilic scalar in the mass range of $\mathcal{O}(10)- \mathcal{O}(1000 )\,\rm GeV$ can be a possible source for the current experimental deviations from the Standard Model (SM) predictions, with $\mathcal{O}(0.1) - \mathcal{O}(1)$ Yukawa couplings. The current electron and muon EDM constraints to the general CP-violating Yukawa couplings are discussed. We propose to search such a leptophilic scalar mediated at the future high-luminosity LHC (HL-LHC) runs, as well as the high-energy lepton colliders, including the CEPC and the muon collider. Our results show that the leptophilic scalar in the mass range of $\mathcal{O}(10)- \mathcal{O}(1000 )\,\rm GeV$ can be fully probed by the future experimental searches at the HL-LHC and the lepton colliders at their early stages.
\end{abstract}

\maketitle


\section{Introduction}
\label{section:intro}

Recently, there have been mild deviations from the precise measurements in the leptonic sector, which include the both muon~\cite{Bennett:2006fi} and electron~\cite{Hanneke:2008tm,Hanneke:2010au} anomalous magnetic moments of
\beqn 
\Delta a_\mu &\equiv&a_\mu^{\rm exp} - a_\mu^{\rm SM}= (274\pm 73) \times 10^{-11}   \,,\label{eq:amu_anomaly}\\
\Delta a_e &\equiv & a_e^{\rm exp} - a_e^{\rm SM} = (-87 \pm 36) \times 10^{-14}\,.\label{eq:ae_anomaly}
\eeqn 
The combined results can be possible indications to the new physics beyond the Standard Model (BSM)~\cite{Czarnecki:2001pv}.
A positive deviation of the muon $g-2$ from the SM prediction of $a_\mu^{\rm SM}=116~591~810(43) \times 10^{-11}$~\cite{Aoyama:2020ynm} has been a long-standing topics for phenomenological studies. 
Meanwhile, the newly measured electron $g-2$, which is a $2.4\,\sigma$ discrepancy from the SM prediction with a negative sign, brings additional challenge to the issue.
In the framework of the supersymmetric models, one has the absolute values of $|\Delta a_\mu| \propto m_\mu^2$ and $| \Delta a_e| \propto m_e^2$.
Therefore, the mismatch of $\Delta a_\mu/ \Delta a_e$ in Eqs.~\eqref{eq:amu_anomaly} \eqref{eq:ae_anomaly} and $| m_\mu / m_e |^2$ disfavors the explanation from the supersymmetric particles.

Among the previous efforts where one tries to address the muon $g-2$ anomaly, it is usually achieved by introducing spin-$0$ and/or spin-$1$ mediators beyond the SM.
 The couplings of these mediators to the leptonic states can be categorized as scalar, pseudo-scalar, vectorial, and axial ones~\footnote{The estimation of the individual contributions to the $g-2$ was made in Ref~\cite{Leveille:1977rc}.}. 
 It is known that the pseudo-scalar and/or axial couplings can lead to negative contributions to the leptonic $g-2$, while the scalar and/or vectorial couplings lead to positive contributions. 
 Therefore, it is meaningful to ask whether one may accommodate both anomalous magnetic moments by minimally including spin-$0$ mediator.
 Through the experience in the SM, or many versions of BSM new physics constructions, it is a general feature that the Yukawa couplings between the scalar fields and the SM fermions can be flavor-dependent.

 Most of the previous studies focus on the simplified model where the scalar or vector mediators are relatively light~\cite{Chen:2015vqy,Marciano:2016yhf,Davoudiasl:2018fbb,Liu:2018xkx,Bauer:2019gfk,Kirpichnikov:2020tcf}, or with the lepton flavor-violating couplings~\cite{Altmannshofer:2016brv,Bauer:2019gfk,Dev:2020drf,Calibbi:2020emz}.
 In addition, there have been systematical studies of both anomalies in the ``full model'', such as in the two-Higgs-doublet model~\cite{Han:2015yys,Wang:2018hnw,Keus:2017ioh,Abe:2017jqo,Han:2018znu,Chun:2019oix,Jana:2020pxx,Botella:2020xzf,Bian:2020vzc,Li:2020dbg,Rose:2020nxm}, the supersymmetric models~\cite{Badziak:2019gaf,Endo:2019bcj}, and others~\cite{Batell:2016ove,CarcamoHernandez:2020pxw,Haba:2020gkr,Keshavarzi:2020bfy,Bigaran:2020jil,Chen:2020jvl}.
Note in passing, the recent efforts in addressing the muon and/or electron $g-2$ anomalies include Refs.~\cite{Crivellin:2018qmi,Crivellin:2020zul,Colangelo:2020lcg}. 
 Meanwhile with the proposals of the future lepton colliders, such as the circular electron-positron collider (CEPC), International Linear Collider (ILC), FCC-ee, and the muon colliders, it is natural to ask whether one can directly look for the corresponding BSM signals.
 Although there can be different configurations for the future $e^+ e^-$ colliders, it is commonly accepted that they will all run at the center-of-mass energy of $\sqrt{s}= 240\, \GeV$ or $\sqrt{s}= 250\,\GeV$ for the precision measurements of the Higgs boson.
 As for the future muon colliders, there has been a renewed interest to look for the physics cases with the center-of-mass energies of $\mO(1) - \mO(100)\,\TeV$~\cite{Buttazzo:2018qqp,Delahaye:2019omf,Costantini:2020stv,Capdevilla:2020qel,Han:2020uid,Long:2020wfp,Han:2020pif,Han:2020uak,Liu:2021jyc}.
 A muon collider which can operate at multiple TeV scale thus provides tremendous physical opportunities beyond that of a Higgs factory operating at several hundred GeV.
 Recently, discussions on the direct probes of the anomalous muon $g-2$ effects at the future muon collider can be also found in Refs.~\cite{Buttazzo:2020eyl,Yin:2020afe}.

With a CP-violating scalar mediator in the leptonic sector, it is natural to expect a prediction to the electric dipole moments (EDM) to the charged leptons.
The most recent upper bounds to the electron and muon EDMs were from the ACME collaboration~\footnote{Another order of magnitude improvement to the electron EDM precision can be expected from the future upgrade.}~\cite{Andreev:2018ayy} and the muon $g-2$ storage ring at the BNL~\cite{Bennett:2008dy} as follows
\beqn\label{eq:EDMbounds}
&& | d_e | < 1.1\times 10^{-29} \, e\cdot {\rm cm} \,, \qquad  | d_\mu | < 1.9\times 10^{-19} \, e\cdot {\rm cm} \,. 
\eeqn 

The rest of the paper is organized as follows.
Our model setup with a CP-violating scalar coupling to the leptonic sector is described in Sec.~\ref{section:model}.
The combined fits to both electron and muon $g-2$ in Eqs.~\eqref{eq:amu_anomaly} and \eqref{eq:ae_anomaly} are performed, with the leptophilic scalar mass in the range of $(10 \,, 1000)\,\GeV$.
The CP-even Yukawa coupling to the muon and the CP-odd Yukawa coupling to the electron will be used as the benchmark model points for the collider searches.
We also impose the electron and muon EDM constraints in Eq.~\eqref{eq:EDMbounds} to the benchmark models with the general CP-violating Yukawa couplings.
In Sec.~\ref{section:LHC}, we study the discovery potentials for the leptophilic scalar mediator at the LHC, within the mass range of $(10\,, 1000)\,\GeV$.
In Sec.~\ref{section:leptonC}, we further study the discovery potentials for the leptophilic scalar mediator at the future high-energy lepton colliders, including both the CEPC running at $\sqrt{s}=240\,\GeV$ and the muon collider running at $\sqrt{s}=3\,\TeV$.
We summarize the discovery potential of the leptophilic scalar $\phi$ at the LHC, CEPC, and the muon collider in Sec.~\ref{section:conclusion}.

\section{The model setup}
\label{section:model}

The minimal model setup involves a CP-violating and leptophilic scalar $\phi$ as follows:
\beqn\label{eq:coup}
\mL&=& \hf (\partial \phi )^2 - \hf m_\phi^2 \phi^2 - \sum_{\ell=e\,, \mu} \bar \ell ( \lambda_\ell  +  \tilde \lambda_\ell   i \gamma^5   )  \ell  \phi   \,.
\eeqn
The presence of the scalar and pseudo-scalar couplings of the $\phi$ field to the SM leptons $\ell$ leads to both magnetic moments and electric dipole moments.
In our setup, we assume the CP-violating Yukawa couplings of $(\lambda_\ell \,, \tilde \lambda_\ell )$ to be flavor-dependent.
Generally speaking, such a CP-violating scalar can induce both the electric and magnetic dipole operators as follows
\beqn 
\mL_{\rm eff}&=& \Big( - \frac{i}{2} d_\ell \bar \ell \sigma^{\mu \nu } \gamma_5 \ell+ \frac{e }{4 m_\ell } a_\ell \bar \ell \sigma^{\mu\nu } \ell  \Big)  F_{\mu\nu } \,,
\eeqn 
through the one-loop radiative corrections.

\subsection{The anomalous $g-2$}

We estimate the size of the anomalous $g-2$ factor of $\Delta a_\ell$ according to the couplings in Eq.~\eqref{eq:coup}.
The one-loop corrections to the lepton $g-2$ from the leptophilic scalar read
\beqn 
\Big( \Delta a_\ell \Big)_{{\rm spin}-0} &=& \lambda_\ell^2 I_S(x) + \tilde \lambda_\ell^2 I_P(x) \,,\label{eq:aell_scalar}\\
I_S(x)&=&\frac{x^2}{ 8\pi^2 } \int_0^1 dz\, \frac{  (1+z)(1-z)^2  }{ x^2 (1-z)^2 + z} \,,\\
I_P(x)&=&-\frac{x^2}{ 8\pi^2 } \int_0^1 dz\, \frac{   (1-z)^3 }{ x^2 (1-z)^2 + z}\,,
\eeqn 
with $x\equiv m_\ell / m_\phi$. 
The integrals of $I_S(x)$ and $I_P(x)$ are for the scalar and pseudoscalar couplings respectively, and their values are opposite.
Therefore, one would expect a CP-violating scalar field coupling to the leptons to be the most economical way.
For a leptophilic scalar mass of $m_\phi \sim \mO(10) - \mO(1000)\,\GeV$, the integrals of $I_{S\,,P}(x)$ are in the range of $\sim \pm \mO(10^{-4}) - \mO(10^{-7})$ for muons and $\sim \pm \mO(10^{-8} ) - \mO(10^{-12} )$ for electrons, respectively.
The absolute values of both integrals vanish with the increasing mediator mass of $m_\phi$.

\begin{figure}[htb]
\centering
\includegraphics[width=0.48\textwidth]{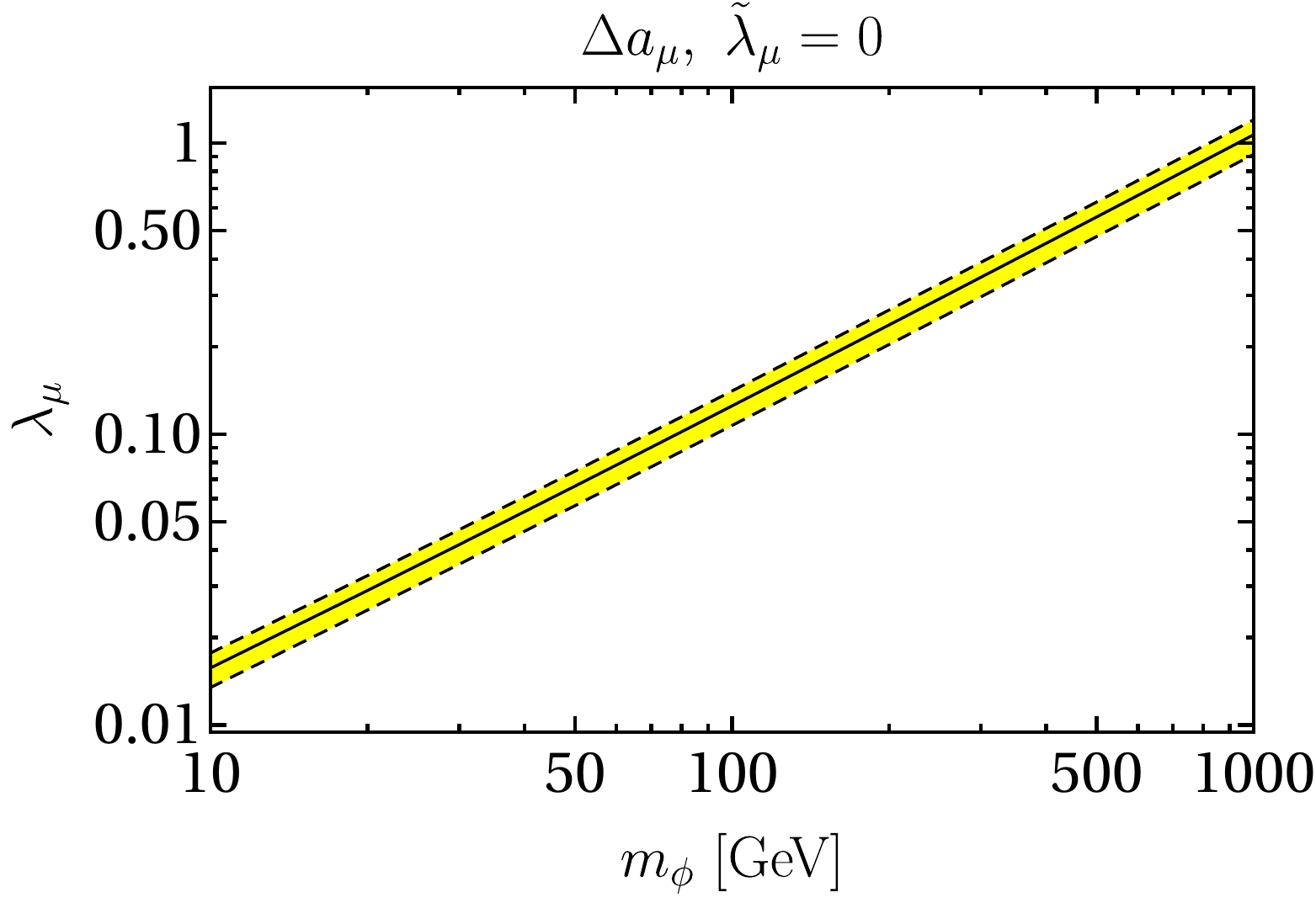}
\includegraphics[width=0.48\textwidth]{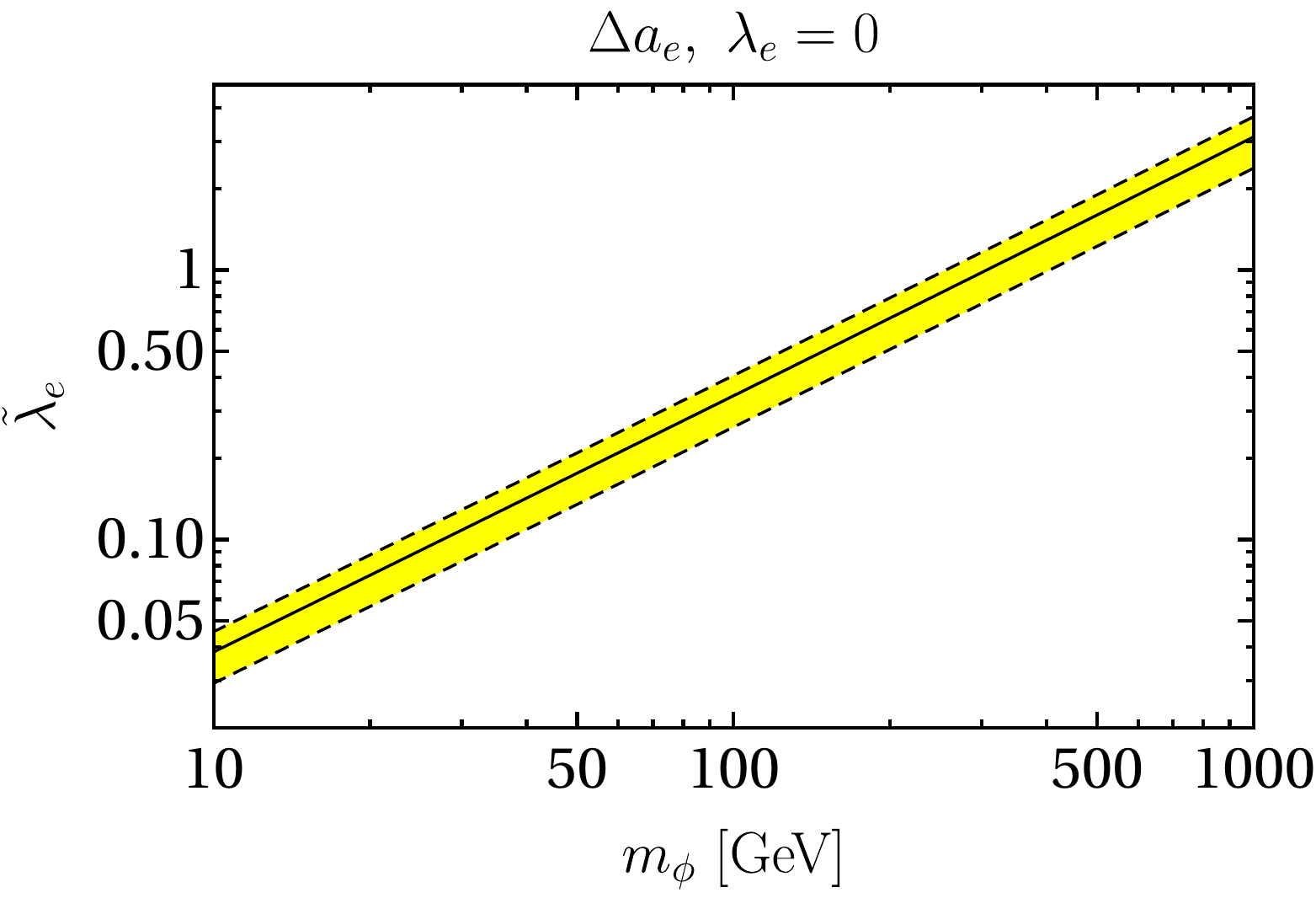}
\caption{\label{fig:g-2}
The $1\,\sigma$ region for the one-loop leptophilic scalar contributions to the muon (left panel) and electron (right panel) $g-2$ in the $(m_\phi\,, \lambda_\mu )$ and $(m_\phi\,, \tilde \lambda_e)$ planes.
}
\end{figure}

To display the one-loop contributions to the muon and electron $g-2$, we vary the scalar mediator mass in the range of $m_{\phi} \in (10\,, 1000)\,\GeV$, and take $\tilde \lambda_\mu = \lambda_e =0$ for simplicity.
By only assuming non-vanishing $\lambda_\mu$ and $\tilde \lambda_e$ in the effective coupling of Eq.~\eqref{eq:coup}, we demonstrate the $1\,\sigma$ region to accommodate the current experimental results of the muon and electron $g-2$ as listed in Eqs.~\eqref{eq:amu_anomaly} and \eqref{eq:ae_anomaly} in Fig.~\ref{fig:g-2}.
With the leptophilic scalar mass increasing from $10\,\GeV$ to $1000\,\GeV$, the corresponding Yukawa couplings of $(\lambda_\mu \,, \tilde \lambda_e)$ can be enhanced from $\mO(0.01)$ to $\mO(1)$.
This implies that for a scalar mediator in such a mass range, the enhancements of the Yukawa couplings bring it possible to probe the scalar mediator at the LHC and the future lepton colliders, such as the CEPC and the muon colliders.

\subsection{The EDM constraints}

In order to accommodate the current $g-2$ measurements, we set $\tilde \lambda_\mu = \lambda_e =0$ in Fig.~\ref{fig:g-2} for simplicity.
In turn, the simultaneous existences of the CP-violating Yukawa couplings lead to the non-vanishing predictions of the lepton EDMs.
The one-loop contribution to the lepton EDM is given by
\beqn\label{eq:lepEDM_OneLoop}
d_\ell &=& \frac{\lambda_\ell \tilde \lambda_\ell }{ 8\pi^2 } \frac{e }{ m_\ell } x^2 \int_0^1 dz \frac{ (1-z)^2 }{ x^2 (1-z)^2 + z}\,.
\eeqn 
Note that it is also possible to induce the lepton EDMs via the two-loop Barr-Zee diagrams.
In practice, we find their contributions to be $\sim \mO(10^{-12})$ smaller than the one-loop contribution in Eq.~\eqref{eq:lepEDM_OneLoop}.

\begin{figure}[htb]
\centering
\includegraphics[width=0.49\textwidth]{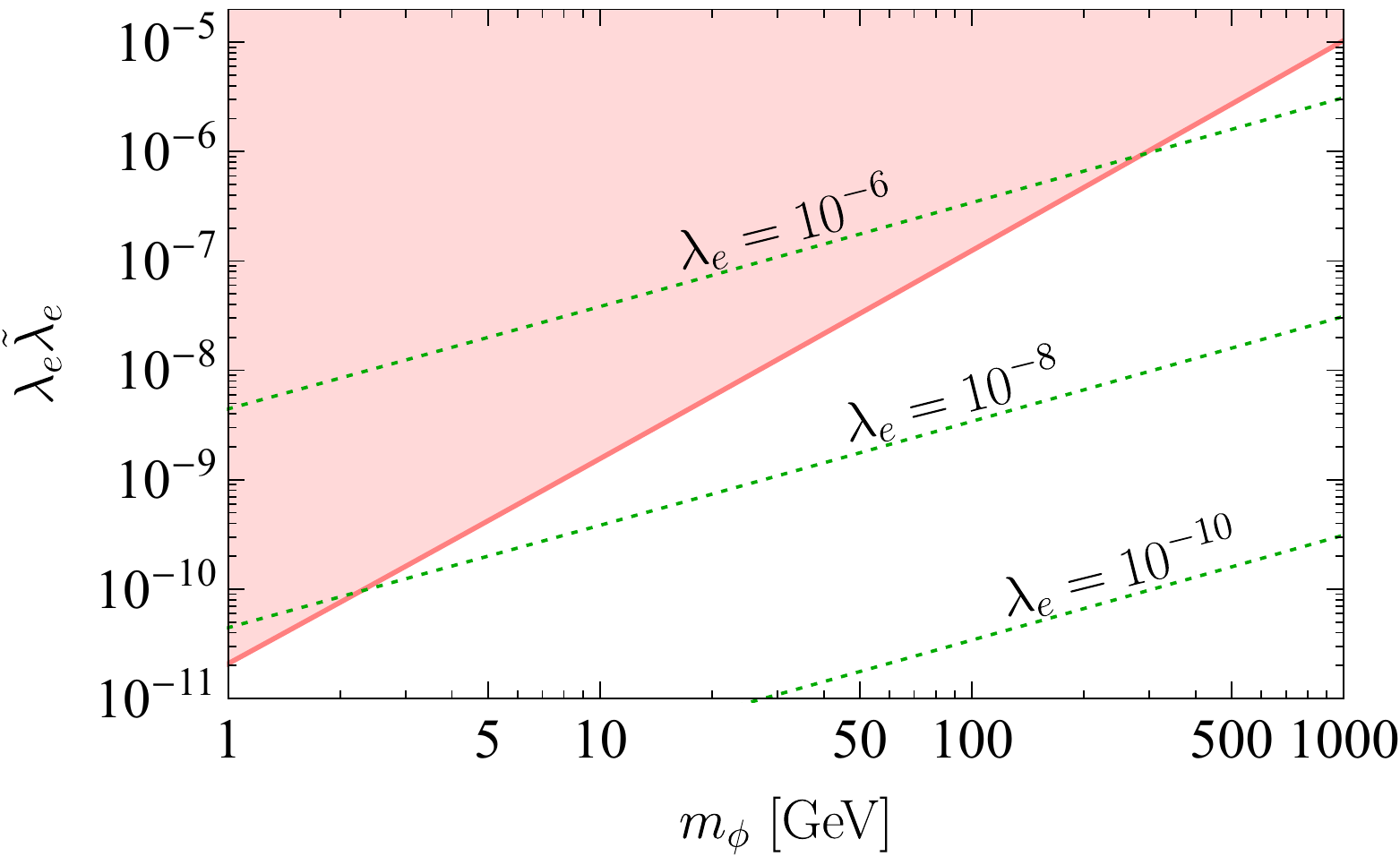}~~
\includegraphics[width=0.48\textwidth]{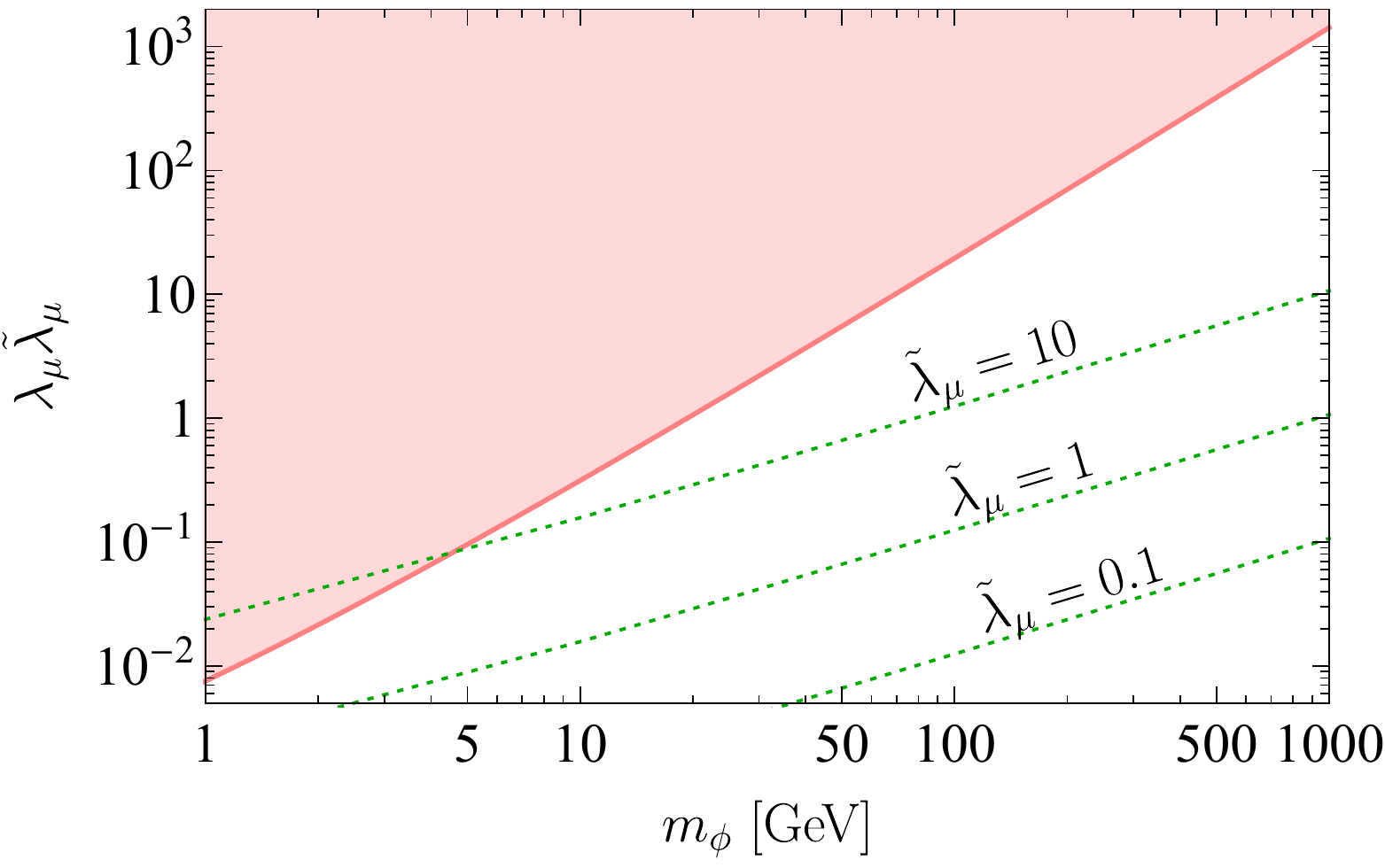}
\caption{\label{fig:EDM}
The bounds to the electron (left panel) and muon (right panel) EDMs in the $(m_\phi\,, \lambda_e \tilde \lambda_e)$ and $(m_\phi\,, \lambda_\mu \tilde \lambda_\mu)$ planes.
The shaded regions represent the electron and muon EDM exclusions according to Eq.~\eqref{eq:EDMbounds}.
}
\end{figure}

According to the one-loop EDM result in Eq.~\eqref{eq:lepEDM_OneLoop}, an upper bound to the lepton EDM imposes constraint to the parameter of $\lambda_\ell \tilde \lambda_\ell$.
In Fig.~\ref{fig:EDM}, we display the electron and muon EDM constraints in the $(m_\phi\,, \lambda_e \tilde \lambda_e)$ and $(m_\phi\,, \lambda_\mu \tilde \lambda_\mu)$ planes.
In the left panel of Fig.~\ref{fig:EDM}, we display three different curves with fixed values of $\lambda_e = (10^{-6} \,, 10^{-8}\,, 10^{-10} )$, and together with the values of $( m_\phi\,, \tilde \lambda_e )$ from the benchmark models as in Fig.~\ref{fig:g-2}.
For the leptophilic scalar in the mass range of $(10\,, 1000)\,\GeV$, the CP-even Yukawa couplings of $\lambda_e \gtrsim 10^{-6}$ have been ruled out by the electron EDM bound. 
In turn, this justifies our previous simplification of $\lambda_e=0$ in fitting the electron $g-2$.
Similarly, in the right panel of Fig.~\ref{fig:EDM} we display three different curves with fixed values of $\tilde \lambda_\mu = (10 \,, 1\,, 0.1 )$, and together with the values of $( m_\phi\,,  \lambda_\mu )$ from the benchmark models as in Fig.~\ref{fig:g-2}.
With the CP-odd Yukawa coupling of $\tilde \lambda_\mu$ in the range of $(0.1\,, 10)$, they can lead to negative contributions to the muon $g-2$.
Since the muon EDM bound is far less stringent comparing to the electron EDM, such parameter ranges cannot be directly excluded.
Thus, we assume that $\tilde \lambda_\mu \ll \lambda_\mu$ in the collider searches to the benchmark models that accommodate the current muon $g-2$ below.


\section{The LHC searches for the leptophilic mediator}
\label{section:LHC}

\subsection{The signal and the background processes at the LHC}

\begin{figure}[htb]
\centering
\includegraphics[width=0.6\textwidth]{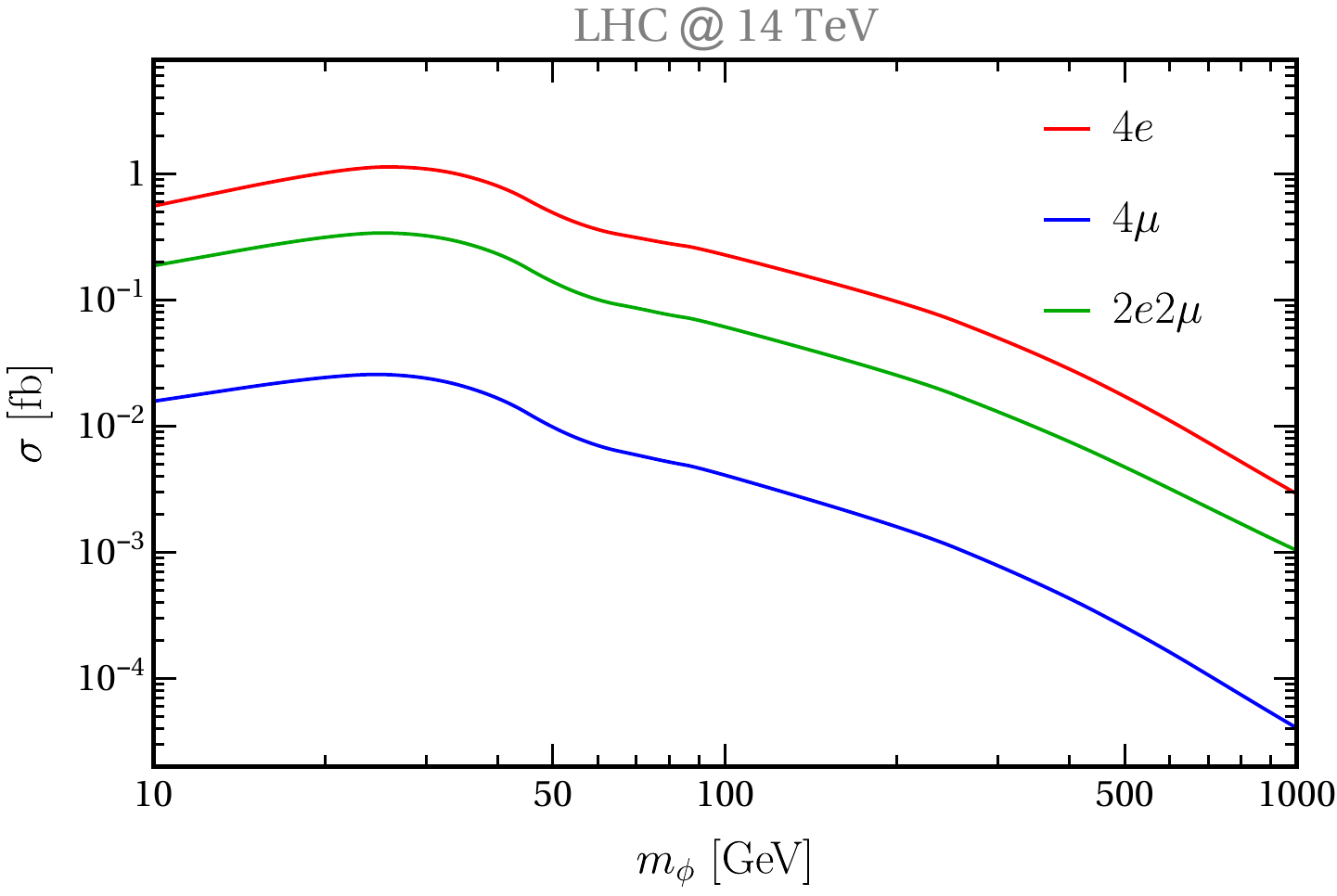}
\caption{\label{fig:LHC_signalxsec}
The signal cross sections of the leptophilic scalar $\phi$ for different final states of $(4e\,, 4\mu \,, 2e 2\mu)$ at the LHC $14\,\TeV$ run.
}
\end{figure}

In this section, we study the future collider searches for the leptophilic scalar $\phi$ at the LHC. 
With the center-of-mass energy of $\sqrt{s} = 14\,\TeV$, and an integrated luminosity of $\mL_{\rm int} = 3000\,\fb^{-1}$, we study the signal processes of
\beqn\label{eq:LHC_sigProcess}
&&  q \bar q \to \gamma^*/Z^* \to \ell^+ \ell^- \phi  ( \to \ell^+ \ell^-  ) \,, 
\eeqn
at the parton level. 
The corresponding cross sections at the LHC $14\,\TeV$ runs are displayed in Fig.~\ref{fig:LHC_signalxsec} for different final states of $(4e\,, 2e 2\mu \,, 4 \mu)$, with $m_\phi \in (10\,, 1000)\,\GeV$.
Among these three different channels, we find that the multiple leptonic final states of $4e$ can have sizable cross sections of $\sim \mO(1)\,\fb$ with $m_\phi \sim \mO(10)\,\GeV$.
While for the $2e 2\mu$ and the $4\mu$ final states, their cross sections are $\sim \mO(10^{-1}) - \mO(10^{-3} )\, \fb$ and $\sim \mO(10^{-2}) -  \mO(10^{-4} ) \,\fb$, respectively.
Hence, we focus on the $4 e$ final states that are most promising at the LHC searches.

The corresponding irreducible SM background processes include~\cite{Sirunyan:2018nnz}
\beqn\label{eq:LHC_bkgdProcess}
&& q \bar q \to e^+ e^-  e^+ e^- \,,\quad  gg \to e^+ e^-  e^+ e^-   \,.
\eeqn
It turns out that the cross sections from the $gg \to 4 e$ contributions are $\sim \mO(1)\,\%$ of the $q \bar q \to 4 e $, and we have the cross section of $\sigma(q \bar q \to e^+ e^-  e^+ e^-)=6.63\, \fb$ at the LHC $14\,\TeV$ run.
Given that the fake rate of muons to electrons can be as low as $\sim \mO(10^{-7})$ at the ATLAS detector~\cite{Aad:2014bca}, we can safely ignore the possible contaminations from the SM background events with the $2e 2\mu$ and/or $4\mu$ final states. 

\subsection{The event generation and selections}

In practice, we generate the signal and background events at the parton level by Madgraph~\cite{Alwall:2011uj}.
Sequentially, they are passed to Pythia8~\cite{Sjostrand:2007gs} for the parton showering and hadronizations, and to Delphes~\cite{deFavereau:2013fsa} for the fast detector simulations.
The events containing four final-state electrons with two opposite-sign pairs are selected. 
Afterwards, four invariant masses from the opposite-sign same-flavor (OSSF) pairs are reconstructed. 
Among them, the one whose invariant mass of $M_{ee}$ is mostly close to the scalar mediator mass of $m_\phi$ is selected.
The significances of $S/\sqrt{B}$ are obtained by imposing the invariant mass cut of $| M_{ ee} - m_\phi | < 0.02\, m_\phi$.

\subsection{The LHC search results}

\begin{figure}[htb]
\centering
\includegraphics[width=0.95\textwidth]{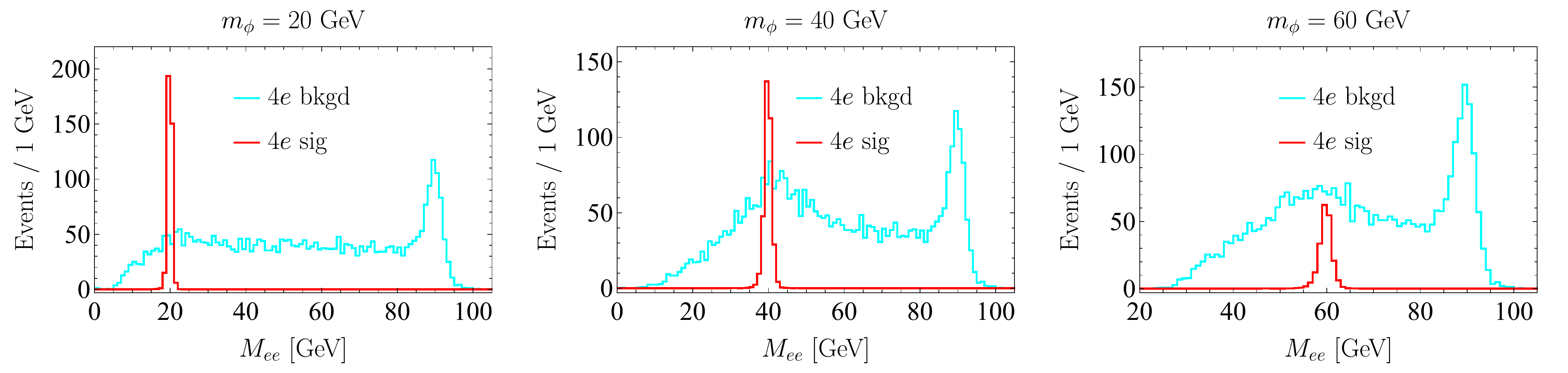}
\caption{\label{fig:LHC_mee}
The invariant mass distributions of $M_{ee}$ for both signals and the SM background events at the LHC $14\,\TeV$ run, with $m_\phi = (20\,, 40\,, 60)\,\GeV$.
}
\end{figure}

We display the invariant mass distributions of $M_{ee}$ in Fig.~\ref{fig:LHC_mee}, with $M_{ee}$ reconstructed from four opposite-sign electrons.
Since the invariant mass of $M_{ee}$ is mostly close to the leptophilic scalar mass of $m_\phi$, the distributions of the corresponding SM background events vary accordingly for different cases.
Here, three different samples with $m_\phi = (20\,, 40\,, 60)\,\GeV$ are displayed. 
One can expect that for the scalar mediator mass close to the $m_Z$, the corresponding distributions of $M_{ee}$ suffer from large SM background events. 
Thus, we do not expect the scalar mediator with mass of $m_\phi \sim 90\,\GeV$ to be discovered at the LHC.
The expected numbers of signal events with integrated luminosity of $3000\,\fb^{-1}$ at LHC  $14\,\TeV$ run are given in Tab.~\ref{tab:cutflow_LHC} for three benchmark models with $m_{\phi}=(20,40,60)\,\GeV$. 
As expected, the SM background is effectively reduced by the mass window cut applied on the reconstructed leptophilic scalar. 
Furthermore, we can see clearly that the significances of $S/\sqrt{B}$ drops rapidly when $m_{\phi}$ get close to $m_Z$.
The integrated luminosities for discovering the leptophilic scalar $\phi$ are to be presented in Fig.~\ref{fig:luminosity}.
It turns out that such leptophilic scalar can be discovered at the HL-LHC run for $m_\phi \in [10\,, 200]\,\GeV$, except for the $Z$-pole region.

\begin{table}[htb]
  \begin{tabular}{|c|c|c|c|c|c|c|c|c|c|}
    \hline\hline
    & \multicolumn{3}{c|}{$m_{\phi}=20~\mathrm{GeV}$} & \multicolumn{3}{c|}{$m_{\phi}=40~\mathrm{GeV}$} & \multicolumn{3}{c|}{$m_{\phi}=60~\mathrm{GeV}$} \\ \cline{2-10}
     & Sig. & Bkgd. & $S/\sqrt{B}$ & Sig. & Bkgd. & $S/\sqrt{B}$ & Sig. & Bkgd. & $S/\sqrt{B}$ \\ \hline
    Select OSSF $4e$ & 367.18 & 3690.68 & 6.04 & 345.20 & 3690.68 & 5.68 & 200.83 & 3690.68 & 3.31 \\
    $|M_{ee}-m_{\phi}|<0.02\,m_{\phi}$ & 216.40 & 34.23 & 36.99 & 216.87 & 117.82 & 19.98 & 133.35 & 173.15 & 10.13 \\
    \hline\hline
  \end{tabular}
  \caption{\label{tab:cutflow_LHC}
  The expected number of events for three benchmark models with $m_{\phi}=(20,40,60)~\GeV$ and the corresponding SM background at the LHC 14 TeV run, with the integrated luminosity of $3000\,\fb^{-1}$.}
\end{table}


\section{The searches for the leptophilic mediator at the lepton colliders}
\label{section:leptonC}

\subsection{The signal and background processes at the lepton colliders}

\begin{figure}[htb]
\centering
\includegraphics[width=0.6\textwidth]{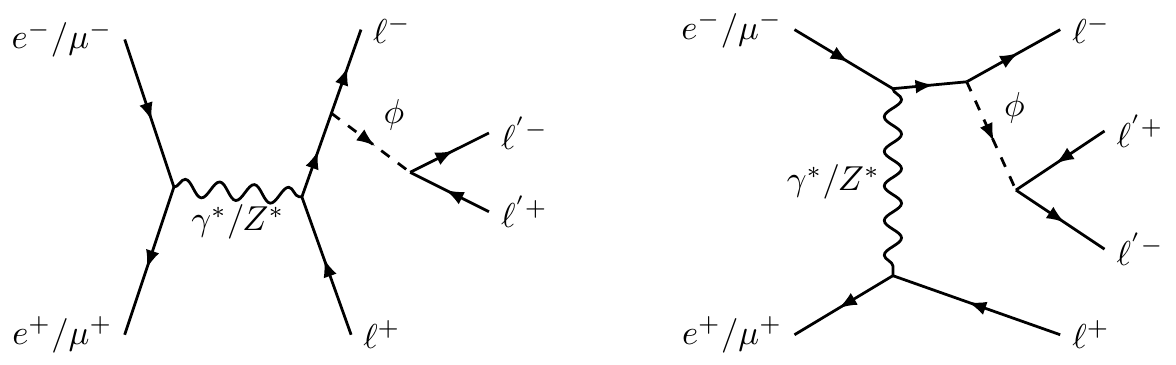}\\
\includegraphics[width=0.9\textwidth]{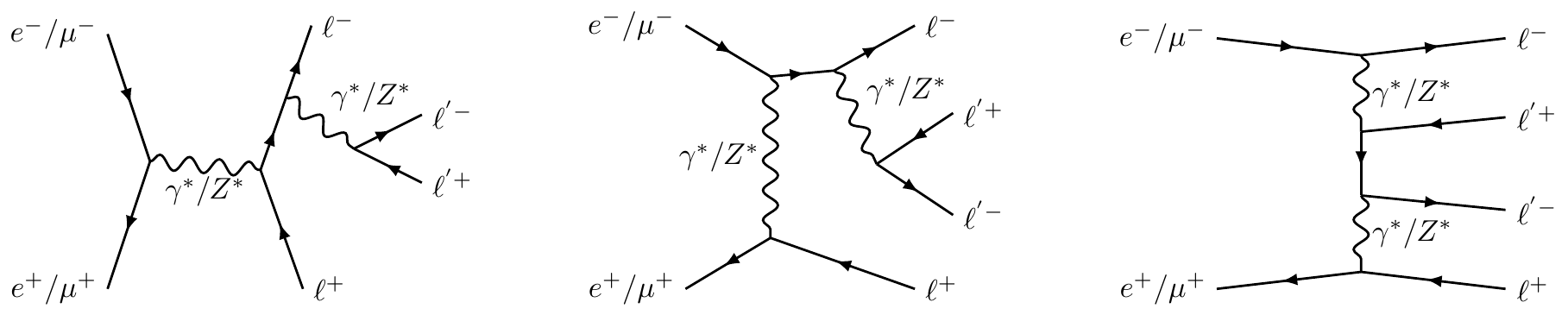}
\caption{\label{fig:CEPC_digrams} 
The Feynman diagrams for the signal (top) and background (bottom) processes at the CEPC and the muon collider runs.
}
\end{figure}

In this section, we proceed to study the future collider searches for the leptophilic scalar $\phi$ at the lepton colliders.
We choose the configurations of the CEPC run at $\sqrt{s} = 240\,\GeV$ and the muon collider run at $\sqrt{s} = 3\,\TeV$.
The following inclusive productions of
\beqn\label{eq:lep_sigProcess}
&& e^+ e^- / \mu^+ \mu^- \to  \phi  \ell^+ \ell^- \to \ell^+ \ell^-  \ell^+ \ell^-  \,,
\eeqn
for the leptophilic scalar $\phi$ will be studied.
The corresponding SM background processes are 
\beqn\label{eq:CEPC_bkgdProcess}
&& e^+ e^-/\mu^+ \mu^- \to  \ell^+  \ell^-  ( \gamma^*/Z) \to  \ell^+ \ell^-  \ell^+ \ell^-   \,,
\eeqn
The related Feynman diagrams are depicted in Fig.~\ref{fig:CEPC_digrams}.
For the sake of simplicity, we only demonstrated few typical diagrams, and the remaining signal or background diagrams can be obtained by attaching the leptophilic scalar $\phi$ or $\gamma^*/Z$ to three other leptons.

\begin{figure}[htb]
\centering
\includegraphics[width=0.48\textwidth]{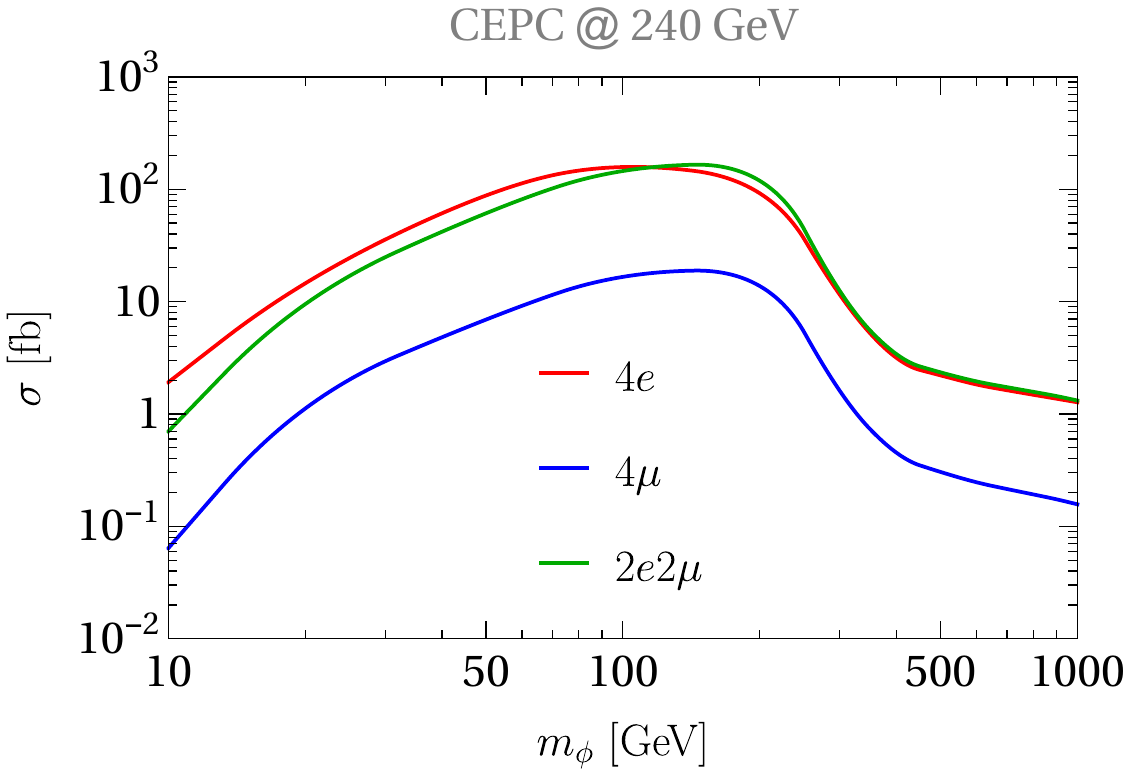}
\includegraphics[width=0.48\textwidth]{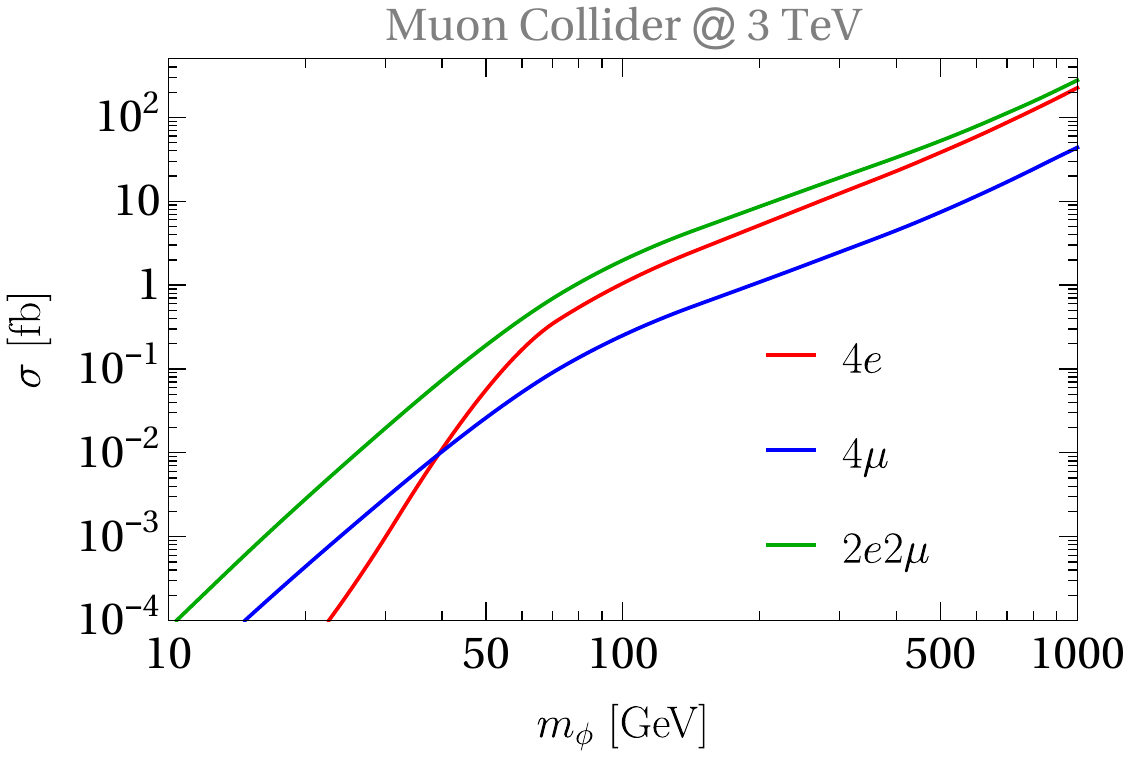}
\caption{\label{fig:lepton_xsec}
The inclusive production cross sections of the leptophilic scalar $\phi$ at the CEPC $240\,\GeV$ run (left panel) and the muon collider $3\,\TeV$ run (right panel), for three different final states of $(4e\,, 4\mu \,, 2e 2\mu )$.
}
\end{figure}

At both the CEPC and the muon collider, the final states of both signal and background events can be categorized into: (i) $4e$, (ii) $4\mu$, and (iii) $2 e 2 \mu $.
The inclusive production cross sections at the CEPC and the muon collider runs are plotted in Fig.~\ref{fig:lepton_xsec} for three different leptonic final states.
At the CEPC $240\,\GeV$ run, the signal cross sections increase from the low-mass range and are peaked around $\sim \mO(100)\,\fb$ for the $4e/ 2e 2\mu$ final states with $m_\phi \sim 100 - 200 \,\GeV$, since the leptophilic scalars are on-shell produced. 
For the leptophilic scalar in the heavy-mass range, the signal cross sections decrease to $\sim \mO(1)\,\fb$ with $m_\phi \sim 1000\,\GeV$ for the $4e/2 e 2 \mu$ final states.
The cross sections for the $4\mu$ final states are smaller compared to the $4e / 2e 2\mu $ cases by an order of magnitude. 
At the muon collider $3\,\TeV$ run, the leptophilic scalar $\phi$ is always on-shell produced.
The production cross sections for three different leptonic final states all increase with the large leptophilic scalar masses of $m_\phi$, and can be $\sim \mO(10) - \mO(100)\,\fb$ when $m_\phi\sim 1000\,\GeV$.
Accordingly, one expects that the muon collider will have good search sensitivities for the leptophilic scalar $\phi$ at the high-mass ranges.

Below, we carry out the analysis of the leptophilic scalar searches at both the CEPC and the muon collider.
The signal and background events are generated at various levels in the same procedures as those in the LHC searches. 
According to the production cross sections, we shall separate the search strategies for the low-mass ranges of $m_\phi \in [50\,, 200]\,\GeV$ at the CEPC, and for the high-mass ranges of $m_\phi \in [200\,, 1000]\,\GeV$ at the muon collider.
Afterwards, the events with two OSSF lepton pairs are selected for each category.
With four selected leptons, we record their kinematic variables of $(p_T\,, \eta\,, \varphi)$ for the reconstructions of the leptophilic scalar.
For the $4e$ and/or $4\mu$ final states, we also order the final-state electrons and/or muons as $(e_1^\pm\,, e_2^\pm)$ and/or $(\mu_1^\pm\,, \mu_2^\pm)$ according to their $p_T$.

\subsection{The CEPC searches}

\begin{figure}[htb]
\centering
\includegraphics[width=0.48\textwidth]{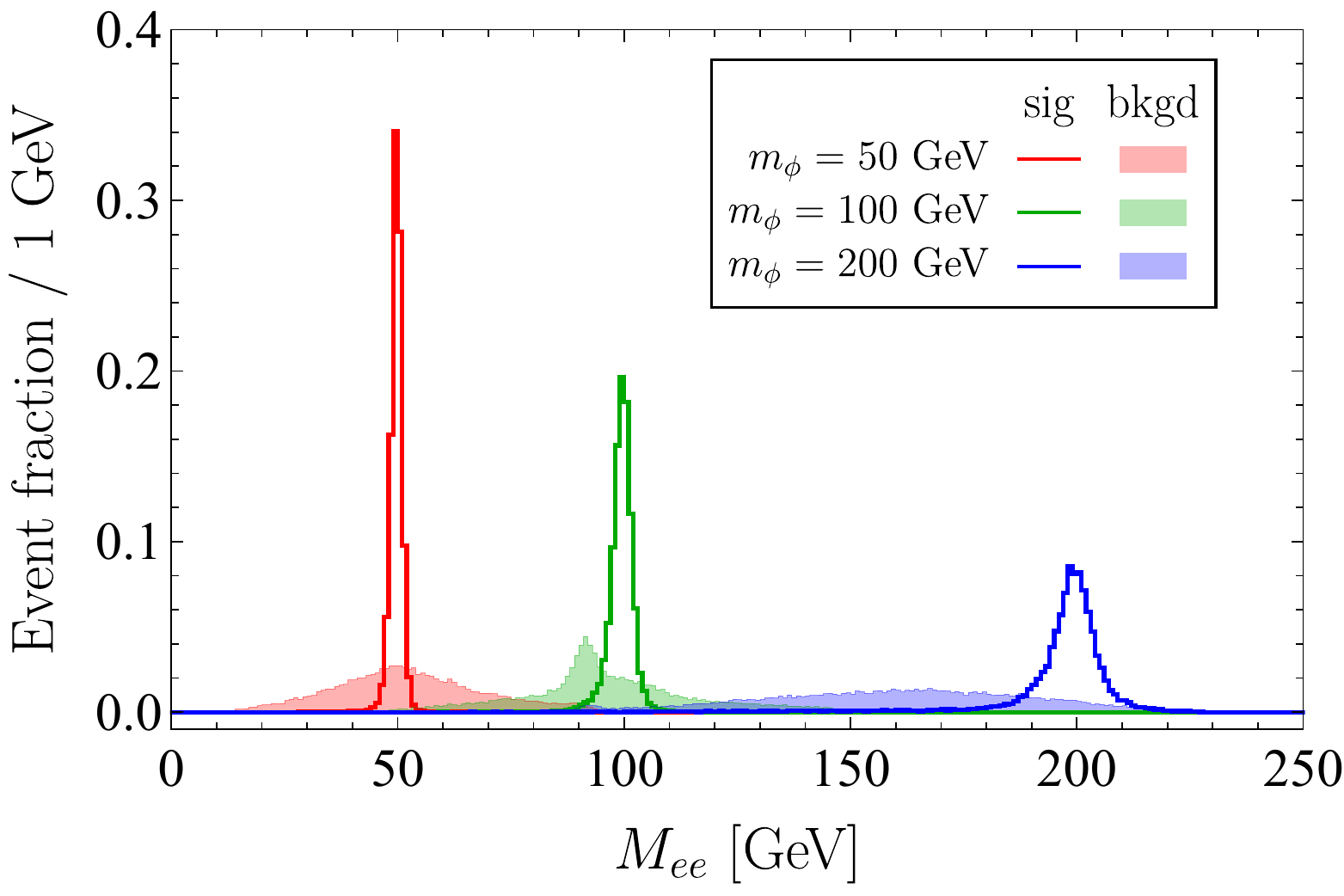}\\
\includegraphics[width=0.48\textwidth]{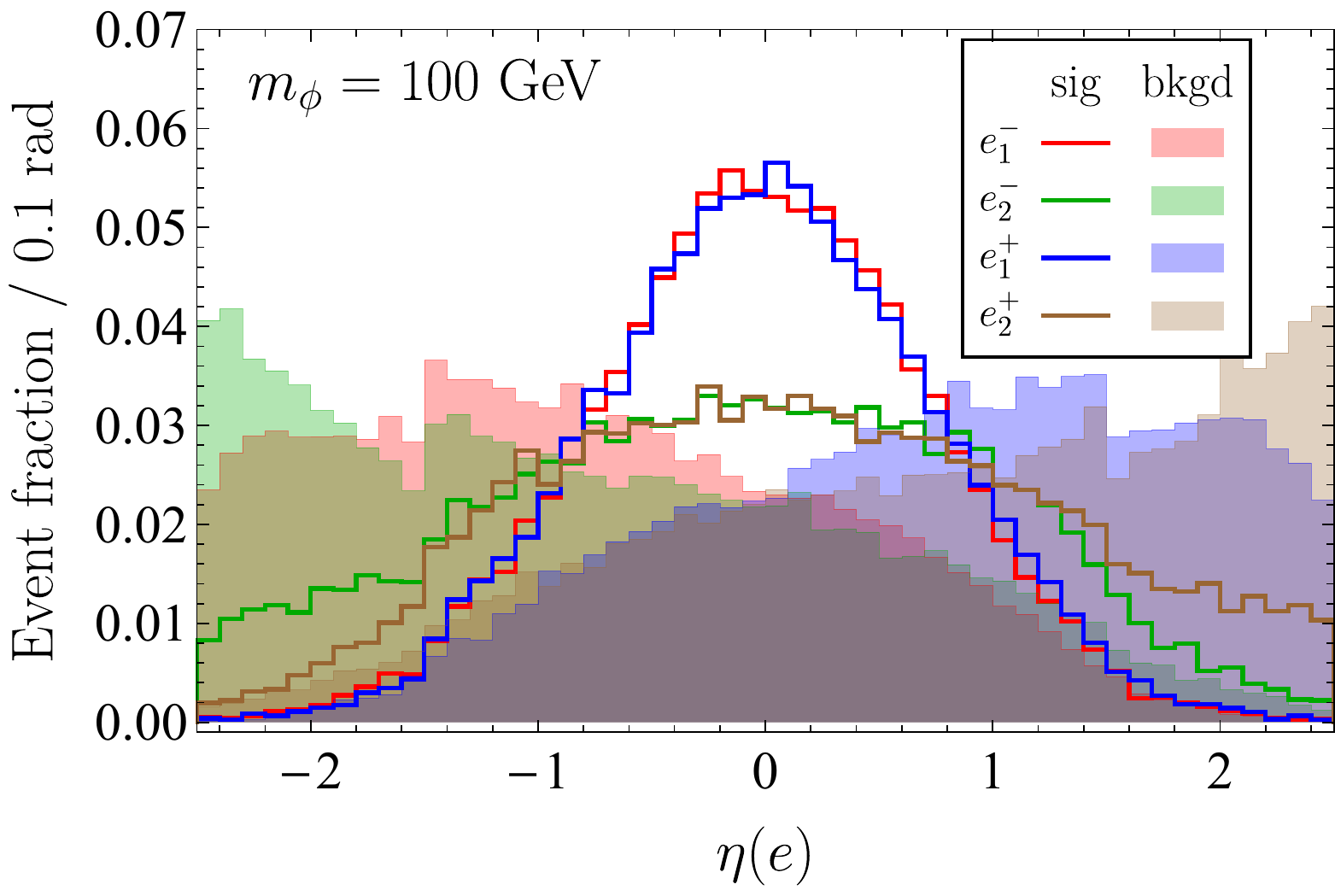}
\includegraphics[width=0.48\textwidth]{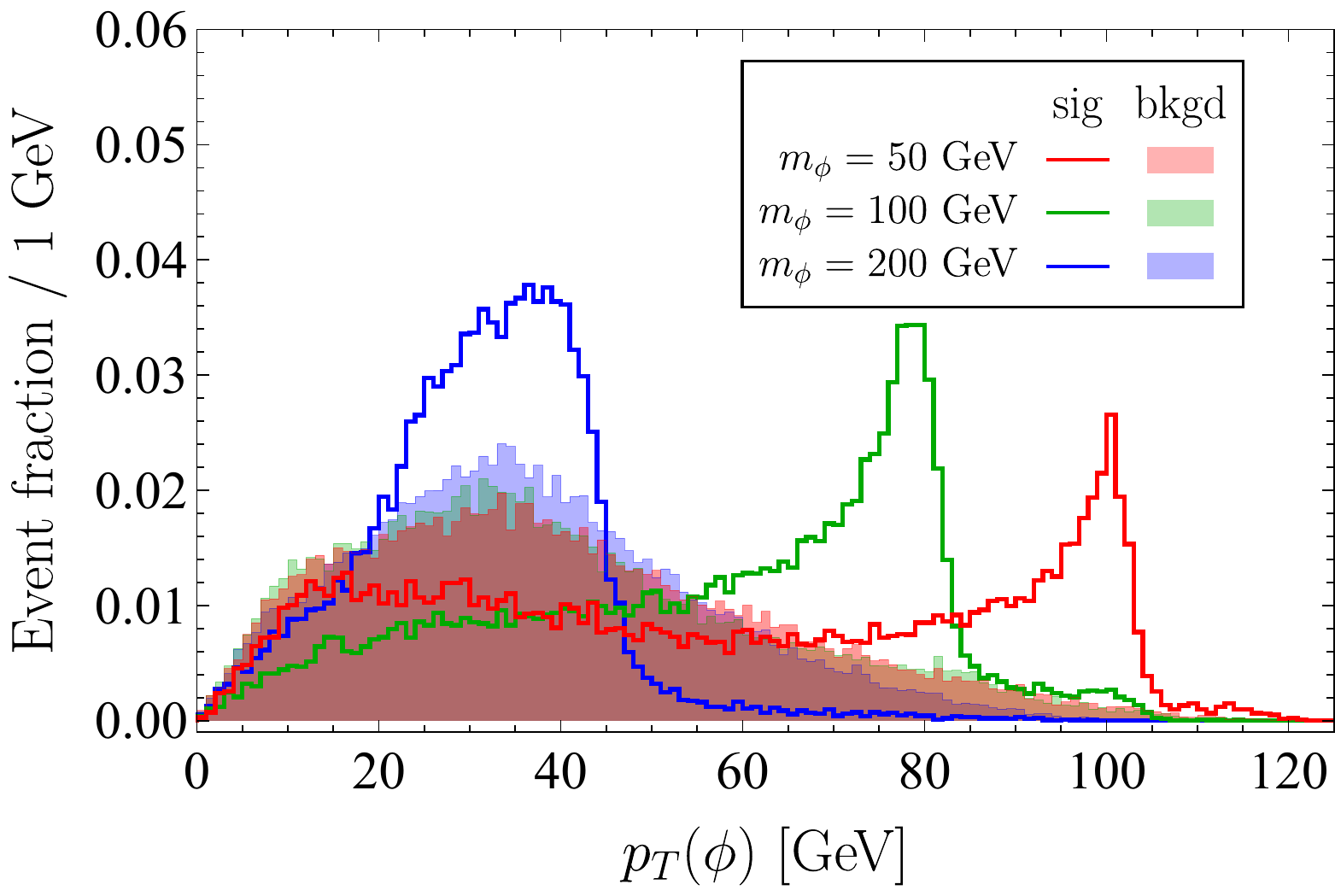}
\caption{\label{fig:CEPC_4edistr}
The distributions of: $\eta(e)$ (bottom left), $p_T(\phi)$ (bottom right), and the invariant mass $M_{ee}$ (top) for both the signal and the SM background events from the $4e$ final states at the CEPC $240\,\GeV$ run.
The $\eta(e)$ distributions are plotted for the $m_\phi=100\,\GeV$ case, while the $p_T(\phi)$ and the $M_{ee}$ distributions are plotted for three different cases of $m_\phi=(50\,, 100\,, 200)\,\GeV$.
}
\end{figure}

At the CEPC $240\,\GeV$ run, one only expects direct $\phi$ resonance productions for $m_\phi \lesssim 240\,\GeV$.
For the low-mass resonances with $m_\phi \in [50\,, 200]\,\GeV$, we obtain two/four invariant masses from the OSSF lepton pairs for the $2e2\mu$/$(4e\,, 4\mu)$ final states.
Among them, the one whose invariant mass of $M_{\ell \ell} (\ell = e\,, \mu)$ is mostly close to the scalar mediator mass of $m_\phi$ is selected.
One can expect this to be the lepton pair from the on-shell productions of the leptophilic scalar, and its $p_T$ and the invariant mass distribution will be used as our kinematic variables to separate the signal events from the SM background events.

In Fig.~\ref{fig:CEPC_4edistr}, we display the event fraction~\footnote{The event fraction is defined such that the summation of the signal and the SM background event fraction is unity.}  distributions of $(\eta(e)\,, p_T(\phi)\,, M_{ee})$ for different samples of $m_\phi=(50\,, 100\,, 200)\,\GeV$ from the $4e$ final states.
For both the $4e$ and $2e 2\mu$ final states, one finds relatively larger values of $|\eta|$ distributions for the final-state electrons in the SM background events.
This is mostly due to the back-to-back productions of the opposite-sign electrons in the background events, as depicted in the last diagram in Fig.~\ref{fig:CEPC_digrams}.
Accordingly, the cuts we choose for the low-mass resonances from the $4e$ and $2e 2\mu$ final states are the following
\begin{itemize}

  \item $m_\phi\in[50,100]\,\GeV$:   $|M_{\ell\ell}-m_\phi|<0.02~m_\phi\,,~p_T(\phi)>60~\text{GeV},~\eta(e^-)>-1,~\eta(e^+)<1$,

  \item $m_\phi=200$~GeV:    $|M_{\ell\ell}-m_\phi|<0.02~m_\phi,\quad p_T(\phi)<60~\text{GeV},\quad \eta(e^-)>-1,\quad \eta(e^+)<1$.

   \end{itemize}
The expected numbers of signal events with integrated luminosity of $1\,\fb^{-1}$ at the CEPC $240\,\GeV$ run are given in Tab.~\ref{tab:cutflow_CEPC} for three benchmark models with $m_{\phi}=(50,100,200)\,\GeV$, respectively. 
We can see that the background events are effectively suppressed by above cuts.
For the $4\mu$ final states, the cuts to the $M_{\ell \ell}$ and $p_T(\phi)$ similarly apply.
However, we do not impose the cuts to the $\eta$ of the final-state muons, since there are no back-to-back productions of muons at the CEPC run.
The final searches sensitivities of the leptophilic scalar $\phi$ via three different final states will be projected in Fig.~\ref{fig:luminosity}.

\begin{table}[htb]
  \begin{tabular}{|c|c|c|c|c|c|c|c|c|c|}
    \hline\hline
    & \multicolumn{3}{c|}{$m_{\phi}=50~\mathrm{GeV}$} & \multicolumn{3}{c|}{$m_{\phi}=100~\mathrm{GeV}$} & \multicolumn{3}{c|}{$m_{\phi}=200~\mathrm{GeV}$} \\ \cline{2-10}
     & Sig. & Bkgd. & $S/\sqrt{B}$ & Sig. & Bkgd. & $S/\sqrt{B}$ & Sig. & Bkgd. & $S/\sqrt{B}$ \\ \hline
    Select OSSF $4e$ & 32.03 & 4.88 & 14.50 & 61.34 & 4.88 & 27.77 & 37.91 & 4.88 & 17.16 \\
    Kinematic Cuts & 7.31 & 0.014 & 61.89 & 14.90 & 0.025 & 93.49 & 15.79 & 0.019 & 115.36 \\
    \hline\hline
  \end{tabular}
  \caption{\label{tab:cutflow_CEPC}
  The expected number of events for three benchmark models with $m_{\phi}=(50,100,200)~\GeV$ and the corresponding SM background at the CEPC 240 GeV run, with the integrated luminosity of $1\,\fb^{-1}$.}
\end{table}

\subsection{The muon collider searches}

\begin{figure}[htb]
\centering
\includegraphics[width=0.48\textwidth]{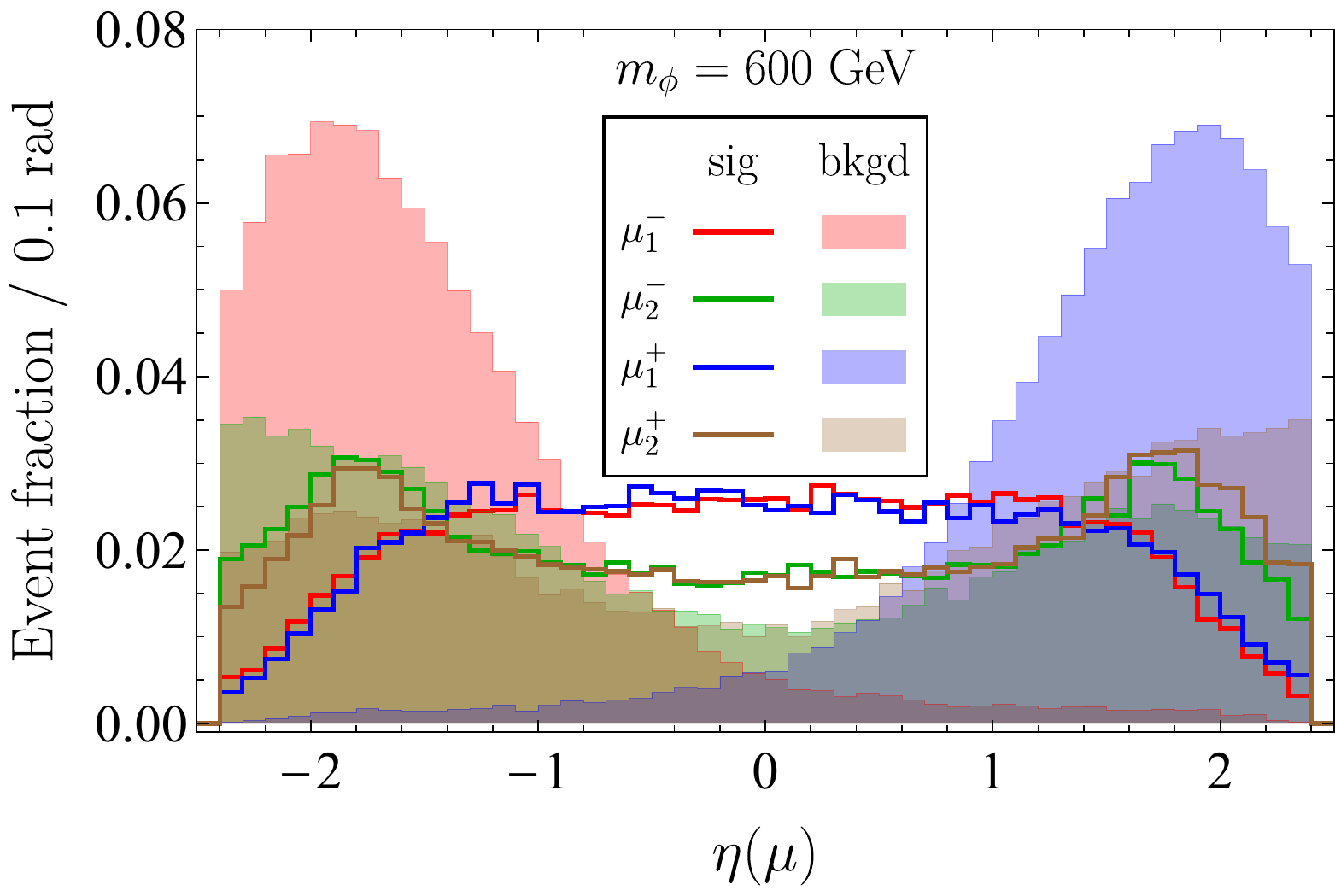}
\includegraphics[width=0.48\textwidth]{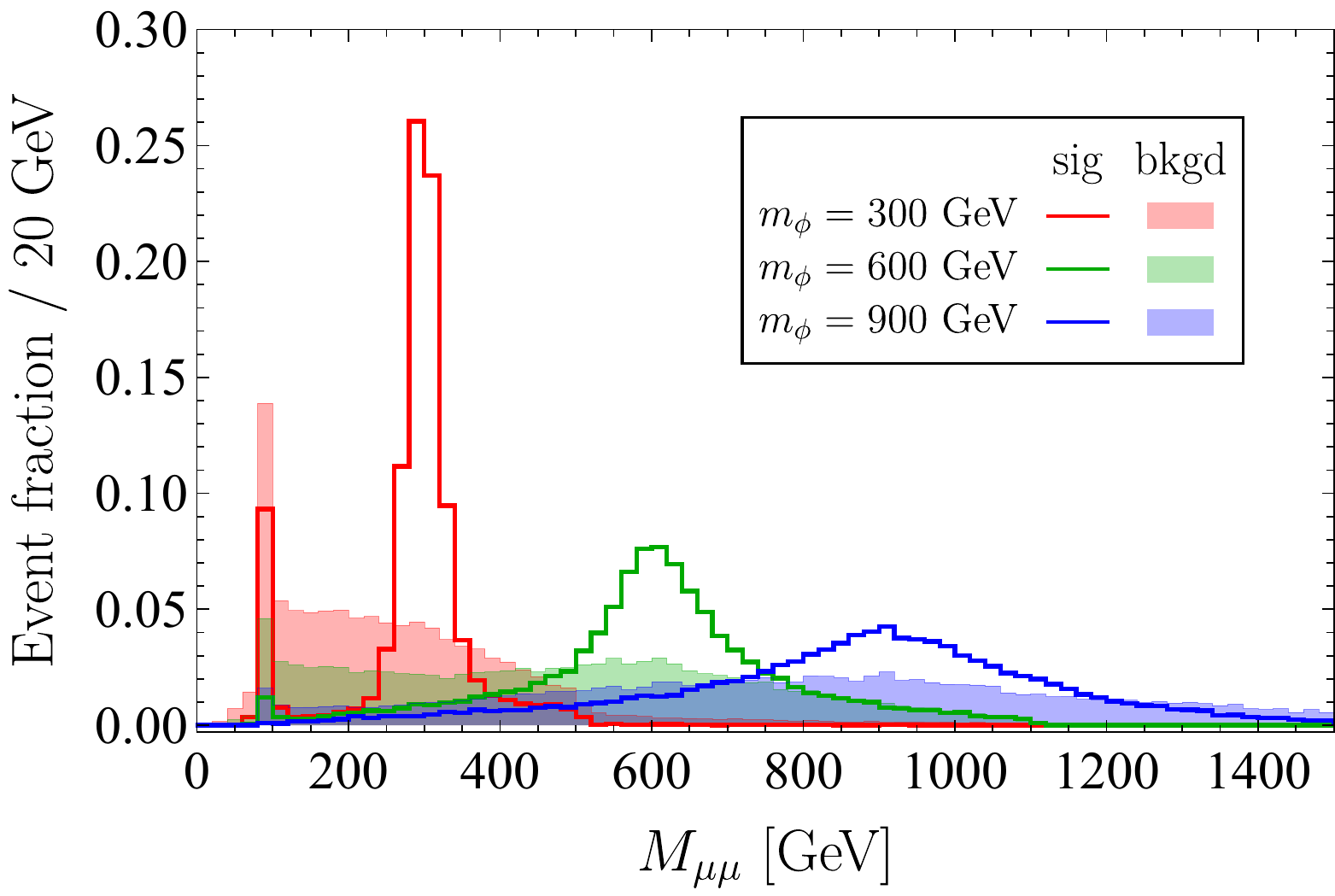}
\caption{\label{fig:muon_4mudistr}
The distributions of: $\eta(\mu)$ (left panel) and the invariant mass $M_{\mu\mu}$ (right panel) for both the signal and the SM background events from the $4e$ final states at the muon collider $3\,\TeV$ run.
The $\eta(\mu)$ distributions are plotted for the $m_\phi=600\,\GeV$ case, while the $M_{\mu\mu}$ distributions are plotted for three different cases of $m_\phi=(300\,, 600\,, 900)\,\GeV$.
}
\end{figure}

Next, we proceed to analyze the searches for the leptophilic scalar $\phi$ at the muon collider $3\,\TeV$ run.
The event selection criteria follow from those in the CEPC case. 
At the muon collider, two/four invariant masses from the OSSF lepton pairs for the $2e 2\mu$/$(4e\,, 4\mu)$ final states can be similarly obtained.
Afterwards, we select the one whose invariant mass of $M_{\ell\ell}\, (\ell = e\,, \mu)$ mostly close to the leptophilic scalar mass of $m_\phi$.
In Fig.~\ref{fig:muon_4mudistr}, we display the event fraction distributions of $ \eta(\mu) $ for $m_\phi =600\,\GeV$, and $M_{\mu\mu}$ for $m_\phi=(300\,, 600\,, 900)\,\GeV$ from the $4\mu$ final states. 
For both the $4\mu$ and $2e 2\mu$ final states at the muon collider, one finds relatively large values of $|\eta|$ distributions for the final-state muons in the SM background events.
At the muon collider here, there can be back-to-back productions of the opposite-sign muons in the background events.
Such a feature does not exist for the $4e$ final states at the muon collider.
However, we find relatively small values of $p_T$ distributions for the final-state $e_1^-$ and $e_1^+$ in the $4e$ signal events.
Accordingly, we choose the following cuts for the high-mass resonances from the final states
\begin{itemize}

  \item $4\mu$ and $2e 2\mu$:  $|M_{\ell\ell}-m_\phi|<0.1\,m_\phi,\quad \eta(\mu_1^-)>0,\quad \eta(\mu_1^+)<0$,

\item $4e$: $|M_{ee}-m_\phi|<0.1\,m_\phi,\quad p_T(e_1^\pm)<900\,{\GeV}$.
   \end{itemize}
The expected numbers of signal events with integrated luminosity of $10\,\fb^{-1}$ at muon collider $3\,\TeV$ run are given in Tab.~\ref{tab:cutflow_Muon} for three benchmark models with $m_{\phi}=(300,600,900)\,\GeV$. 
We can see that the background events are effectively suppressed by above cuts.
The final search sensitivities of the leptophilic scalar $\phi$ via three different final states will be projected in Fig.~\ref{fig:luminosity}.

\begin{table}[htb]
  \begin{tabular}{|c|c|c|c|c|c|c|c|c|c|}
    \hline\hline
    & \multicolumn{3}{c|}{$m_{\phi}=300~\mathrm{GeV}$} & \multicolumn{3}{c|}{$m_{\phi}=600~\mathrm{GeV}$} & \multicolumn{3}{c|}{$m_{\phi}=900~\mathrm{GeV}$} \\ \cline{2-10}
     & Sig. & Bkgd. & $S/\sqrt{B}$ & Sig. & Bkgd. & $S/\sqrt{B}$ & Sig. & Bkgd. & $S/\sqrt{B}$ \\ \hline
    Select OSSF $4\mu$ & 14.80 & 4.59 & 6.91 & 71.78 & 4.59 & 33.51 & 210.41 & 4.59 & 98.24  \\
    Kinematic Cuts & 3.08 & 0.011 & 29.31 & 8.97 & 0.018 & 66.25 & 19.65 & 0.024 & 127.57 \\
    \hline\hline
  \end{tabular}
  \caption{\label{tab:cutflow_Muon}
  The expected number of events for three benchmark models with $m_{\phi}=(300,600,900)~\GeV$ and the corresponding SM background at the muon collider 3 TeV run, with the integrated luminosity of $10\,\fb^{-1}$.}
\end{table}


\section{Conclusion and Discussion}
\label{section:conclusion}

\begin{figure}[htb]
\centering
\includegraphics[width=0.7\textwidth]{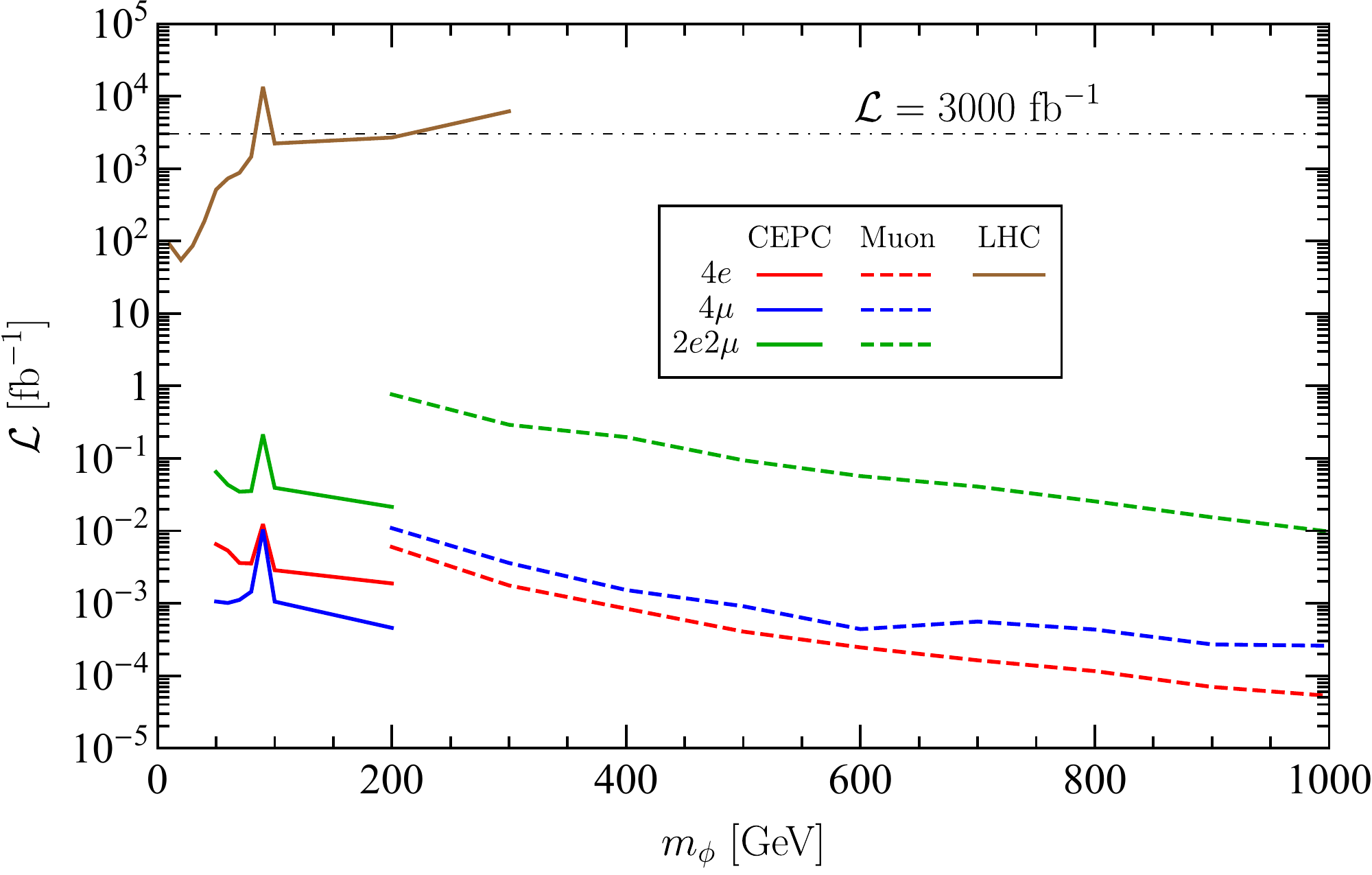}
\caption{\label{fig:luminosity} 
The combined discovery potentials of the leptophilic scalar $\phi$ at the LHC $14\,\TeV$ run (brown solid curve), CEPC $240\,\GeV$ run (solid curves), and the muon collider $3\,\TeV$ run (dashed curves).
}
\end{figure}

In this work, we suggest a leptophilic scalar with CP-violating Yukawa couplings so that the current discrepancies of the electron and muon $g-2$ can be accommodated.
In comparison to the previous efforts, we find that the reasonable Yukawa couplings can be $\sim \mO(0.1)- \mO(1)$ when the scalar mass is $\sim \mO(10) - \mO(1000)\,\GeV$.
With the sizable Yukawa couplings in this mass range, it is expected that such a leptophilic scalar can be searched for at the current LHC experiments, as well as the future high-energy lepton colliders, such as the CEPC and the muon colliders running at the $\sqrt{s} \sim \mO(1)\,\TeV$.
The related collider search strategy can be realized straightforwardly by constructing the OSSF final-state lepton pairs and impose the invariant mass cuts accordingly.

Our final results of searching for such a leptophilic scalar $\phi$ are summarized in Fig.~\ref{fig:luminosity} at three different collider runs.
The benchmark models with the leptophilic scalar $\phi$ in the mass range of $(10\,, 1000)\,\GeV$ have been studied, where the integrated luminosities needed for the $5\,\sigma$ discovery at the LHC $14\,\TeV$ run, the CEPC $240\,\GeV$ run, and the muon collider $3\,\TeV$ run are displayed.
We find that the HL-LHC run with the integrated luminosity of $3000\,\fb^{-1}$ can probe the leptophilic scalar $\phi$ in the mass range of $[10\,, 200]\,\GeV$ via the $4e$ final state, except for the $Z$-pole region.
The CEPC run can probe such a leptophilic scalar via all three $(4e \,, 4\mu \,, 2e 2\mu)$ final states with integrated luminosities of $\sim \mO(10^{-3}) - \mO(10^{-1})\,\fb$, including the $Z$-pole region.
For the leptophilic scalar in the high-mass range of $[200\,, 1000]\,\GeV$, we find that the muon collider run can fully probe the benchmark models at early stage with integrated luminosities of $\sim \mO(10^{-4}) - \mO(1) \,\fb^{-1}$.


\section*{ACKNOWLEDGMENTS}

We would like to thank Yanwen Liu for very useful discussions and communication during the preparation of this work. 
NC and BW are partially supported by the National Natural Science Foundation of China (under Grant No. 11575176 and 12035008).
CYY is supported in part by the National Natural Science Foundation of China (under Grant No.~11975130, 12035008, 12047533), by the National Key Research and Development Program of China (under Grant No.~2017YFA0402200), and the China Post-doctoral Science Foundation (under Grant No.~2018M641621).

\newpage


\appendix


%

\end{document}